\documentstyle[a4,12pt,amscd]{amsart}
\def\CP#1{\Bbb{P}^{#1}}

\def\SetRe{\Bbb{R}}
\def\SetCx{\Bbb{C}}

\def\sfrac#1#2{{\textstyle\frac{#1}{#2}}}
\def\CY{Calabi-Yau}
\def\K{{K\"ahler}}

\def\ob{\frak{A}}

\def\pD{\cal{D}}
\def\mm{\cal{M}}
\def\W{\cal{W}}
\def\V{\cal{V}}
\def\nlsm{non-linear $\sigma$-model}

\def\opeq#1{\advance\lineskip#1 \advance\baselineskip#1
	\advance\lineskiplimit#1}
\def\eqalign#1{\null\,\vcenter{\opeq{2.5\jot}\mathsurround=0pt
	\everycr={}\tabskip=0pt
	\halign{\strut\hfil$\displaystyle{##}$&$\displaystyle{{}##}$\hfil
	\crcr#1\crcr}}\,\null}

\newcommand{\B}{\cal B}
\newcommand{\E}{\cal E}
\newcommand{\F}{\cal F}
\renewcommand{\O}{\cal O}
\renewcommand{\P}{\Bbb P}
\newcommand{\Image}{\operatorname{Image}}
\newcommand{\supp}{\operatorname{supp}}

\title{Topological Field Theory and Rational Curves}
\author[P.S. Aspinwall]{Paul S. Aspinwall}
\address{University of Oxford\\
Department of Theoretical Physics\\
1 Keble Road\\
Oxford OX1 3NP, U.K.}
\author[D.R. Morrison]{David R. Morrison}
\address{Department of Mathematics\\
Duke University\\
Durham\\
NC 27706, U.S.A.}
\date{October 1991}

\begin{document}
\maketitle
\setlength{\unitlength}{1in}

\hfill
\begin{picture}(1.1,0)
\put(0,2.2){\makebox{\parbox{1.1in}{{\sc outp-91-32p} \\
                                    {\sc duk-m-91-12} }}}
\end{picture}

\begin{abstract}
  We analyze the quantum field theory corresponding to a string
propagating on a \CY\ threefold. This theory naturally leads to the
consideration of Witten's topological non-linear $\sigma$-model and
the structure of rational curves on the \CY\ manifold. We study in
detail the case of the world-sheet of the string being mapped to a
multiple cover of an isolated rational curve and we show that a
natural compactification of the moduli space of such a multiple cover
leads to a formula in agreement with a conjecture by Candelas,
de la Ossa, Green and Parkes.
\end{abstract}

\section{Introduction}

In its most fundamental form, string theory is normally considered as
a loop of string propagating through space-time to sweep out a
two-dimensional world-sheet. This map of the world-sheet into
space-time allows one to ``pull back'' physics from the familiar
space-time around us into a more simple two-dimensional quantum field
theory. This model is usually cast in the form of the \nlsm\ where
the non-linearity of this field theory arises
from curvature in the target space. Thus it is a simple matter to
solve string theory in flat space-time but more general curved target
spaces can only be solved perturbatively assuming the curvature is
small in some sense.

When one specializes to the case of requiring space-time supersymmetry
one picks out a specific class of allowed target spaces in the above
approximation, namely that the target space should have vanishing
Ricci curvature and be a complex K\"ahler manifold. When building
realistic models of physics one is naturally led in this situation to
considering Calabi-Yau threefolds~\cite{CHSW:,Cal:sm}.

This model might have been of interest only to superstring
phenomenologists if it weren't for the fact that there is an
alternative to this method of solution. That is, one can take an
algebraic approach to string theory as a conformal field
theory~\cite{BPZ:,FMS:}. This process does not involve the
approximations required in the \nlsm.
The analogue of the Calabi-Yau condition in
this case is that the conformal field theory should be an N=2
superconformal field theory. At first it was naturally assumed that
this provided a more general class of solutions to the superstring
than \CY's ever could but it is now generally believed from evidence
such as~\cite{Gep:} that these two approaches are in some way
equivalent although this equivalence can only be manifest in some
regions of the moduli space of the theories considered~\cite{AL:qag}.

This equivalence should be a considerable source of new results as the
knowledge to date about \nlsm s and N=2 superconformal field theories
is quite different. Thus, for example, a great deal of knowledge about
the classification of superconformal field theories should be obtained
from the classification of \CY\ manifolds which is better understood.
Going the other way, the mirror symmetry which is trivial in the
language of conformal field theories has led to a new symmetry
hitherto unsuspected in algebraic geometry~\cite{GP:mir,ALR:coup}.
Combining the knowledge of the \nlsm\ with the mirror symmetry has
also allowed further analysis of the structure of a particular
\CY\ and has allowed information about the rational curves on this manifold
to be obtained in an elegant way~\cite{Cand:rat}. It is the purpose of
this paper to analyze the link between the rational curves on a
\CY\ manifold and another quantum field theory closely related to the
superconformal field theory, namely the topological quantum field
theory.

The topological quantum field theory was first introduced in the context
of Yang-Mills theory by Witten~\cite{W:tqft} to provide a model of
Donaldson's diffeomorphism invariants for 4-manifolds~\cite{Don:,Don:poly}.
Witten then introduced a topological version of the
\nlsm~\cite{W:tsm} which will be our main object of study in this
paper. This will allow us to build diffeomorphism invariant
polynomials with integer coefficients for \CY\ manifolds. The integer
coefficients are related to the number of rational curves within a
certain class (a generalization of degree) when this number is finite.

In section 2 we study the N=2 \nlsm\ and show how it is related to the
topological $\sigma$-model. In section 3 we begin the calculation of
the integer coefficients reducing it to a problem in algebraic
geometry. In section 4 we perform this calculation and finally in
section 5 we combine the results.


\section{Reduction to Topological Field Theory}

Let us consider a map of a Riemann surface $\Sigma$ into a manifold
$X$
\begin{equation}
  \phi:\Sigma\rightarrow X.
\end{equation}
Putting coordinates $u^\mu$ on $X$ and $\sigma^\alpha$ on $\Sigma$ we
can locally consider this map to be determined by
functions $u(\sigma)$ on $\Sigma$.
The bosonic string action (see for example~\cite{Cal:sm}) then
associates the following \nlsm\ to this
\begin{equation}
  S=t\int d^2\sigma(\eta^{\alpha\beta}g_{\mu\nu}\partial_\alpha u^\mu
	\partial_\beta u^\nu +
	\varepsilon^{\alpha\beta}B_{\mu\nu}\partial_\alpha u^\mu
	\partial_\beta u^\nu). \label{eq:bnlsm}
\end{equation}
where we have imposed a flat (at least locally) metric
$\eta^{\alpha\beta}$ on $\Sigma$ and $\varepsilon^{\alpha\beta}$ is the
antisymmetric tensor. The symmetric field $g_{\mu\nu}$ is a function of $u^\mu$
and represents the metric on $X$ and there is an antisymmetric field
$B_{\mu\nu}$, also a function of $u^\mu$. The field $B_{\mu\nu}$ does
not have as clear a geometric interpretation as does $g_{\mu\nu}$.
The real parameter, $t$, can be thought of as representing the string
tension or, equivalently, it can be absorbed into $g_{\mu\nu}$ and
$B_{\mu\nu}$ and so used to control the volume of $X$.

If we now impose N=2 supersymmetry we are led
to the condition that $X$ is a complex \K\ manifold~\cite{Zum:}.
We will assume that this manifold has $h^{2,0}=0$ so that all
2-forms are (1,1)-forms (up to addition of exact forms).
Replacing $\sigma^\alpha$ with the single complex coordinate $z$ and
replacing the real index structure of $u^\mu$ with a \K\ index
structure we obtain the bosonic part of the action from (\ref{eq:bnlsm})
\begin{equation}
  S=\frac t2\int\left\{
	g_{i\bar\jmath}(\partial u^i\bar\partial u^{\bar\jmath} +
	\bar\partial u^i\partial u^{\bar\jmath}) +
	iB_{i\bar\jmath}(\partial u^i\bar\partial u^{\bar\jmath} -
	\bar\partial u^i\partial u^{\bar\jmath})\right\}d^2z.
\end{equation}
We can split this as follows
\begin{equation}
  \eqalign{
  S&=t\int g_{i\bar\jmath}\bar\partial u^i\partial u^{\bar\jmath}\,d^2z +
	\frac t2\int(g_{i\bar\jmath} +
	iB_{i\bar\jmath})(\partial u^i\bar\partial u^{\bar\jmath} -
	\bar\partial u^i\partial u^{\bar\jmath})\,d^2z \cr
   &=t\int g_{i\bar\jmath}\bar\partial u^i\partial u^{\bar\jmath}\,d^2z
	+ \frac t2\int_\Sigma(\phi^*J).} \label{eq:snlsm}
\end{equation}
where $\phi^*J$ is the pullback
from $X$ to $\Sigma$ of a (1,1)-form whose real part is the usual
\K-form on $X$.

Let us now add in fermions (i.e., anticommuting operators) $\chi^i$
and $\rho^i$ and another bosonic field $F^i$
along with their conjugates, $\chi^{\bar\imath}$,
$\rho^{\bar\imath}$ and $F^{\bar\imath}$, to obtain an explicitly N=2
supersymmetric action. The above field theory splits naturally into
holomorphic (left-moving) and anti-holomorphic (right-moving) pieces
so that we can obtain 2 sets of N=2 supersymmetries generated by the
infinitesimal parameters $\epsilon_n,\bar\epsilon_n$. We have
\begin{xalignat}{2}
  \delta_{\epsilon_1} u^i&=i\epsilon_1\chi^i,
	 &\delta_{\bar\epsilon_1} u^i&=0, \notag\\
	\delta_{\epsilon_1} u^{\bar\imath}&=0,
	 &\delta_{\bar\epsilon_1}
		u^{\bar\imath}&=i\bar\epsilon_1\chi^{\bar\imath},\notag\\
	\delta_{\epsilon_1} \chi^i &= 0,
	 &\delta_{\bar\epsilon_1} \chi^i&=0, \notag\\
	\delta_{\epsilon_1} \chi^{\bar\imath} &= 0,
	 &\delta_{\bar\epsilon_1} \chi^{\bar\imath}&=0, \notag\\
	\delta_{\epsilon_1} \rho^i&=i\epsilon_1F^i,
	 &\delta_{\bar\epsilon_1} \rho^i&=
		2\bar\epsilon_1\bar\partial u^i,\notag\\
	\delta_{\epsilon_1} \rho^{\bar\imath}&=
		2\epsilon_1\partial u^{\bar\imath},
	 &\delta_{\bar\epsilon_1} \rho^{\bar\imath}&=
		i\bar\epsilon_1F^{\bar\imath},\notag\\
	\delta_{\epsilon_1} F^i&=0,
	 &\delta_{\bar\epsilon_1} F^i&=-2\bar\epsilon_1
		\bar\partial\chi^i,\notag\\
	\delta_{\epsilon_1} F^{\bar\imath}&=-2\epsilon_1
		\partial\chi^{\bar\imath},
	 &\delta_{\bar\epsilon_1} F^{\bar\imath}&=0, \label{eq:N=2(1)}
\end{xalignat}
and
\begin{xalignat}{2}
  \delta_{\epsilon_2} u^i&=i\epsilon_2\rho^i,
	 &\delta_{\bar\epsilon_2} u^i&=0, \notag\\
	\delta_{\epsilon_2} u^{\bar\imath}&=0,
	 &\delta_{\bar\epsilon_2}
		u^{\bar\imath}&=i\bar\epsilon_2\rho^{\bar\imath},\notag\\
	\delta_{\epsilon_2} \chi^i&=-i\epsilon_2F^i,
	 &\delta_{\bar\epsilon_2} \chi^i&=
		2\bar\epsilon_2\partial u^i,\notag\\
	\delta_{\epsilon_2} \chi^{\bar\imath}&=
		2\epsilon_2\bar\partial u^{\bar\imath},
	 &\delta_{\bar\epsilon_2} \chi^{\bar\imath}&=
		-i\bar\epsilon_2F^{\bar\imath}, \notag\\
	\delta_{\epsilon_2} \rho^i &= 0,
	 &\delta_{\bar\epsilon_2} \rho^i&=0, \notag\\
	\delta_{\epsilon_2} \rho^{\bar\imath} &= 0,
	 &\delta_{\bar\epsilon_2} \rho^{\bar\imath}&=0,\notag\\
	\delta_{\epsilon_2} F^i&=0,
	 &\delta_{\bar\epsilon_2} F^i&=2\bar\epsilon_2
		\partial\rho^i,\notag\\
	\delta_{\epsilon_2} F^{\bar\imath}&=2\epsilon_2
		\bar\partial\rho^{\bar\imath},
	 &\delta_{\bar\epsilon_2} F^{\bar\imath}&=0, \label{eq:N=2(2)}
\end{xalignat}
We can now write down the N=2 supersymmetric form
of (\ref{eq:snlsm}) with the fermions $\chi$ and $\rho$. The fields
$F^i$ can be determined through the equations of motion to yield
\begin{equation}
  F^i=-\Gamma^i_{jk}\chi^i\rho^k,\quad\quad
  F^{\bar\imath}=-\Gamma^{\bar\imath}_{\bar\jmath\bar k}
	\chi^{\bar\jmath}\rho^{\bar k}
\end{equation}
where $\Gamma^i_{jk}$ is the usual connection on a \K\ manifold,
$g^{i\bar m}\partial_jg_{k\bar m}$, (it does not depend on
$B_{i\bar\jmath}$). Eliminating the $F^i$ fields we have
\begin{equation}
\begin{split}
  S=t\int\Bigl\{g_{i\bar\jmath}\bar\partial u^i\partial u^{\bar\jmath}
	&- \sfrac i2g_{i\bar\jmath}\rho^iD\chi^{\bar\jmath}
	- \sfrac i2g_{i\bar\jmath}\rho^{\bar\jmath}\bar D\chi^i
	- \sfrac14R_{i\bar\imath j\bar\jmath}\chi^i\chi^{\bar\imath}
	\rho^j\rho^{\bar\jmath}\Bigr\}d^2z \\
	&+ \frac t2\int_\Sigma(\phi^*J).
\end{split} \label{eq:nlsm}
\end{equation}
In this equation, $D$ is the covariant derivative, e.g., $\bar D\chi^i
= \bar\partial\chi^i+\Gamma^i_{jk}\bar\partial u^j\chi^k$ and
$R_{i\bar\imath j\bar\jmath}=\partial_i\partial_{\bar\jmath}
g_{j\bar\imath}-g_{m\bar m}\Gamma^m_{ij}
\Gamma^{\bar m}_{\bar\imath\bar\jmath}$.
The above action is invariant (the integrand is unchanged
up to total derivatives) under (\ref{eq:N=2(1)}) and
(\ref{eq:N=2(2)}) if $g_{i\bar\jmath}$ is a \K\ metric and $J$ is a
closed 2-form.

Consider the last term in (\ref{eq:nlsm}) which we shall refer to as $S_c$.
The first thing one should notice is that this term has not
changed from (\ref{eq:snlsm}), that is, it has the rather strange
property of being supersymmetric even though it does not explicitly
contain any fermions. The $S_c$ term is also a topological
invariant, i.e., it depends only on the cohomology class of $J$ and
the homology class of the image of $\Sigma$ in $X$.
Under small continuous variations of the fields $(u^i,\chi^i,\rho^i)$
$S_c$ does not change.
Thus we can say
that, in some sense, this term has no bosonic degrees of freedom and no
fermionic degrees of freedom. This does not stop this term from being
of some importance however.

The first integral, $S_t$, in (\ref{eq:nlsm}) does not have this
property. That is, it is not topological in the above sense. It is,
however, topological in another sense. It is the precise action
(except that holomorphic and antiholomorphic maps are interchanged) that
appears as a topological \nlsm\ in~\cite{W:tsm}. Also the part of the
N=2 supersymmetry given by (\ref{eq:N=2(1)}) is the BRST symmetry used
in that theory. Note that we have to reinterpret the r\^oles of the
fields $\chi$ and $\rho$ when going between the language of N=2
theories and topological theories. That is, we introduced $\chi$ and
$\rho$ as world-sheet spinors but to interpret $S_t$ as a topological
field theory we consider $\chi^i$ to be a 0-form and $\rho^i$ to be a
(0,1)-form on $\Sigma$. This reinterpretation is usually referred to
as ``twisting'' the N=2 theory.

There are two ingredients in establishing a topological quantum field
theory. Firstly one must identify a BRST symmetry, $Q$. This
automatically exists in an N=2 theory and is provided by
(\ref{eq:N=2(1)}). Secondly, the action is itself BRST-exact (i.e., it is
the BRST variation of something).
This is true of $S_t$ but not of $S_c$:
\begin{equation}
  \delta_{\epsilon_1}\delta_{\bar\epsilon_1}\Bigl\{-\frac t4
	\int g_{i\bar\jmath}\rho^i\rho^{\bar\jmath}\,d^2z
	\Bigr\}=\epsilon_1\bar\epsilon_1S_t.
\end{equation}
A theory with
these two properties will have correlation functions which are integers
(with the observables in a suitable basis)
and does not depend on any continuous parameters in the theory, i.e.,
the $n$-point functions are diffeomorphism invariant~\cite{W:tqft}.
Thus our action for an N=2 \nlsm\ does not have any interesting
topological properties as a whole but it splits nicely into two pieces
$S_c$ and $S_t$ which each have special properties.

Consider calculating a 3-point function between 3 observables
$\ob_a$
\begin{equation}
  \eqalign{
  f_{abc} &=\langle\ob_a\ob_b\ob_c\rangle\cr
  &= \int\ob_a\ob_b\ob_ce^{-S}\,\pD u\pD\chi\pD\rho.\cr}
\end{equation}
Because $S_c$ is invariant under smooth continuous
deformations of the fields, it
is simplest to consider this path integral as a discrete sum over the
homotopy classes of the map $\phi$. This is the usual instanton
calculation. For classes $v$ we have
\begin{equation}
  f_{abc}=\sum_v f^{(v)}_{abc}
\end{equation}
where
\begin{equation}
  \eqalign{f^{(v)}_{abc}&=\int_v\ob_a\ob_b\ob_ce^{-S}\,
	\pD u\pD\chi\pD\rho\cr
	&=e^{-S_c(v)}\int_v\ob_a\ob_b\ob_ce^{-S_t}\,
	\pD u\pD\chi\pD\rho.\cr}
\end{equation}
This shows that the only path integral we need perform is within the
topological field theory. Strictly speaking, we have a different
stress-energy tensor between the N=2 superconformal field theory and
the topological model but this does not affect the calculation for
3-point functions~\cite{W:aspects}.

As usual in instanton calculations, we take the approximation of
expanding around the classical solutions of the theory. For
topological field theories this is not an approximation at all since
we know that changing continuous parameters in the theory such as $t$
in (\ref{eq:nlsm}) have no effect on the correlation functions. Thus
we can take the $t\to\infty$ limit suppressing all but the $S_t=0$
contributions to the path integral. The whole theory is not invariant
however and so one should be careful to realize that we
are always assuming that we are in
some large $t$, i.e., large radius limit in the following
calculations.
The topological action is of the form
\begin{equation}
  S_t=t\int d^2z\Big\{\|\bar\partial u^i\|^2 + \hbox{fermions}\Big\}
\end{equation}
and so we see that the instantons are given by the {\em holomorphic\/}
maps $u^i(z)$. This leads to three possibilities for the map $\phi$:
\begin{enumerate}
  \item $\phi(\Sigma)$ is a point in $X$.
  \item $\phi(\Sigma)$ is an algebraic curve in $X$.
  \item $\phi(\Sigma)$ is a multiple cover of an algebraic curve in
	$X$.
\end{enumerate}
We shall often refer to all of the above cases as $n$-fold covers
of algebraic curves with $n=0$ for case
(1)\footnote{The target point does not need to be contained in such a curve
in $X$ in this case.}.

Let $e_a, a=1\ldots h^{1,1}$ be a set of integral generators of
$H^{1,1}(X)$.
We therefore have
\begin{equation}
  [J] = 2\sum_{a=1}^{h^{1,1}}c^a[e_a]
\end{equation}
where $c^a$ are complex numbers. It follows that
\begin{equation}
  S_c = t\sum_{a=1}^{h^{1,1}}c^am_a \label{eq:scm}
\end{equation}
where $m_a$ are integers given by the homology class of
$\phi(\Sigma)$. The case $n=0$ comprises of all $m_a$ vanishing.
Let us introduce a new set of variables
\begin{equation}
  q_a = e^{-tc^a},
\end{equation}
so that
\begin{equation}
  e^{-S_c} = q_1^{m_1}q_2^{m_2}\ldots q_{h^{1,1}}^{m_{h^{1,1}}}.
\end{equation}
The observables, $\ob_a$, in the topological field theory correspond to
cohomology classes on $X$~\cite{W:tsm}. The N=2 superconformal field
theory has 2 sets of fields that naturally correspond to the
(1,1)-forms and the (2,1)-forms on $X$.
When an N=2 theory is twisted to form a topological field theory, the
process of introducing a BRST symmetry forces us to project out one of
these sets of fields. That is, observables must be BRST-closed and
this can only be achieved for one of the classes of fields at a time.
In this case, we are left with
only the (1,1)-forms as observables. One can twist the N=2 model in a
different way however and be left with only the (2,1)-forms. We will
have more to say about this later. We can take, as observables,
the generators $e_a$
considered as elements of $H^{1,1}(X)$ or, equivalently by Poincar\'e
duality, as algebraic surfaces within $X$ to form 3-point functions
$f_{abc}$ which will have the general form
\begin{equation}
  f_{abc}=N^0_{abc} + N^1_{abc}q_1 + N^2_{abc}q_2 + \ldots +
	N^{\ldots}_{abc}q_1^2q_2 + \ldots  \label{eq:fab}
\end{equation}
That is, these $f_{abc}$ functions can be expressed as polynomials
in $q_i$. The coefficients of the polynomial, $N^v_{abc}$, are the
3-point functions from the topological field theory and, as we will
see in the next section, these numbers are intersection numbers of
cycles in a moduli space and, in particular, are integers.

Thus we see that to each $X$, the \nlsm\ associates a set of
polynomials $f_{abc}$ in $h^{1,1}$ variables, $q_i$. The $t\to\infty$
limit, implicit in calculating $f_{abc}$, corresponds to limit
$q_i\to0$. If we were to calculate the 3-point functions for the
purely topological field theory, i.e., if we drop the $S_c$ term from
the action, we would lose the $q_i$ dependence of $f_{abc}$ (as one
would expect for a topological field theory). In fact, dropping the
$S_c$ term corresponds to setting $q_i=1$ which in turn is like a
``zero radius limit'' of (\ref{eq:fab}), that is, we shrink $X$ down
to zero so that $J=0$ in (\ref{eq:nlsm}). We thus have the rather
curious result that the topological field theory (or, at least its
3-point functions) related to the usual
string action is obtained by taking the large radius approximation of
the \nlsm\ and then putting the radius equal to zero!

It is important to note that we will not explicitly use any
information about the complex structure of $X$ in deriving $f_{abc}$.
These functions depend explicitly only on the cohomology class of $J$.
The $f_{abc}$ are thus local diffeomorphism invariant objects of $X$
although they could conceivably depend on global changes in complex
structure in much the same way as Donaldson's polynomials~\cite{Don:poly}.

In the large radius limit, $f_{abc}$ is dominated by the $N^0$ term in
(\ref{eq:fab}). This integer, corresponding to the case where
$\phi(\Sigma)$ is a point, is given by the intersection number of the
4-cycles $e_a,e_b,e_c$ within the threefold $X$~\cite{W:aspects}. It is
useful to think of the functions $f_{abc}$ as generalizations of the
intersection form to contain more diffeomorphism invariant information.

Having reduced the functions $f_{abc}$ to polynomials with integer
coefficients we will now go on to analyze the precise meaning of these
integers as regards the geometry of $X$.


\section{Reduction to Algebraic Geometry}

We need to compute
\begin{equation}
  N^v_{abc} = \int_ve_ae_be_c\,\exp(-S_t)\,\pD u\pD\chi\pD\rho,
		\label{eq:11c}
\end{equation}
where
\begin{equation}
  S_t=t\int\Bigl\{g_{i\bar\jmath}\bar\partial u^i\partial u^{\bar\jmath}
	- \sfrac i2g_{i\bar\jmath}\rho^iD\chi^{\bar\jmath}
	- \sfrac i2g_{i\bar\jmath}\rho^{\bar\jmath}\bar D\chi^i
	- \sfrac14R_{i\bar\imath j\bar\jmath}\chi^i\chi^{\bar\imath}
	\rho^j\rho^{\bar\jmath}\Bigr\}d^2z. \label{eq:11a}
\end{equation}
The index structure of the fields $\chi^i$ and $\rho^i$ should be
interpreted as these fields taking values in $R=\phi^*T_X$,
the pullback of the holomorphic tangent bundle on $X$, with the barred
quantities taking values in the pullback of the antiholomorphic tangent
bundle. The field $u^i$ however takes values in $X$ itself rather than
the tangent bundle.
Any small variation, $\delta u^i$, can be considered as a deformation
of $X$ and so this field does take values in the pullback of the
tangent bundle. In particular, it will be important to note that the
field $\bar\partial u^i$ is an $R$-valued (0,1)-form.

The integral (\ref{eq:11c}) was effectively solved some time ago for
the case where $\phi(\Sigma)$ is a point~\cite{SW:coup} and when
$\phi(\Sigma)$ is a rational curve~\cite{DSWW:np}. In the latter case
the idea is that
one assumes the $t\to\infty$ limit and so restricts the integral to
the case $\bar\partial u^i=0$, i.e., the holomorphic maps. One
interprets the path integral as being over a {\em superspace\/}
parametrized by $(u^i,\chi^i,\rho^i)$. The $\bar\partial u^i=0$
constraint restricts this to the moduli superspace comprising of the
moduli space of holomorphic maps in question and the fermion zero
modes.
To yield the value of (\ref{eq:11c}) one integrates over this moduli
superspace.

The BRST invariance gives a grading (a BRST ``charge'') to the fields
in the theory. The charges can be assigned as follows;
$q(u^i)=0$, $q(\chi^i)=1$,
$q(\rho^i)=-1$ and $q(e_a)=1$. The path integration measure also has a
BRST charge which comes from the fermion zero modes. Each $\chi^i$ zero
mode contributes $-1$ and each $\rho^i$ zero mode contributes $+1$. It
is also known that the fermion zero modes correspond to harmonic forms
on $\Sigma$. Therefore
\begin{align}
  \hbox{\# of $\chi$ zero modes}&=\dim H^0(R)\notag\\
  \hbox{\# of $\rho$ zero modes}&=\dim H^1(R).
\end{align}
The 3-point function (\ref{eq:11c}) vanishes unless the BRST charges
cancel. This leads to the constraint
\begin{equation}
  \dim H^0(R)-\dim H^1(R)=3.
\end{equation}
The left hand side of this equation is the index of the $R$-twisted
Dolbeault complex on $\Sigma$ and can easily be
computed (see \cite{CKM:hdcg}) by the index theorem
(or the Hirzebruch-Riemann-Roch theorem):
\begin{equation}
  \eqalign{
  \dim H^0(R)-\dim H^1(R)&=\int_\Sigma\hbox{ch}(R).\hbox{td}(T_\Sigma)\cr
   &= (1-g)\dim X - \deg(\phi^*K_X),\cr}
\end{equation}
where $g$ is the genus of $\Sigma$ and $K_X$ is the canonical divisor
of $X$. Thus we have the result that for a \CY\ threefold, we will
generically have non-zero values for $f^v_{abc}$ for all classes of
maps $v$ of genus 0 curves $\Sigma$ into $X$.

Let us now specialize to the case where $X$ is a \CY\ threefold and we
will specify $v$ as being a class of maps from $\Sigma$, a genus 0
Riemann surface, to give an $n$-fold cover of an
isolated rational curve, $C$, in $X$. We will assume that the curve is
of type ($-1$,$-1$), i.e.,
\begin{equation}
  T_X|_C\cong \O_C(2)\oplus\O_C(-1)\oplus\O_C(-1).
\end{equation}
Thus
\begin{equation}
  \eqalign{
  \dim H^0(R) &= \dim H^0(\O(2n)\oplus\O(-n)\oplus\O(-n))\cr
	&= 2n+1\cr}
\end{equation}
and
\begin{equation}
  \eqalign{
  \dim H^1(R) &= \dim H^0(K_\Sigma\otimes R^{-1})\cr
	&=\dim H^0(\O(-2-2n)\oplus\O(n-2)\oplus\O(n-2))\cr
	&=2n-2.\cr}
\end{equation}

If $n=1$, life is comparatively simple. There are no $\rho$ zero modes
and three $\chi$ zero modes. Putting these into (\ref{eq:11c}), the three
operators for the observables $e_a$ ``soak up''
the $\chi$ zero modes and we can do the
integral over the moduli space $Sl(2,\SetCx)$ without considering the
$\exp(-S_t)$ term that appears in (\ref{eq:11c}). (See~\cite{DG:exact}
for the explicit calculation.) When $n>1$, things become more messy.
We now have $\rho$ zero modes and an equal number of extra $\chi$ zero
modes. These cannot all be soaked up by the observables. The integral is
not zero however since we can expand out the $\exp(-S_t)$ term to
introduce some $R_{i\bar\imath j\bar\jmath}\chi^i\chi^{\bar\imath}
\rho^j\rho^{\bar\jmath}$ terms into the action. These terms soak up
the extra fermion zero modes that have appeared to yield a
generically non-zero
answer. Such a calculation will be, in general, rather difficult.

There is a much more elegant approach however exploiting the fact that
we are dealing with a topological field theory. A topological field
theory is based on an action which has the general form
\begin{equation}
  S_t=t\int d^2z\Big\{\|s\|^2 + \hbox{fermions}\Big\} \label{eq:act1}
\end{equation}
where $s$ is a section of a bundle $\W$ over some moduli
space, $\mm$, of all field configurations. When calculating a path integral
containing the above action, a topological field theory has the
property that such an integral will not depend on the parameter $t$.
One can thus take the limit $t\to\infty$ so that the path integral
only contains contributions from the zero locus of $s$. What's more,
the fermionic part of the action is arranged so that the determinants
arising from the integral over the fermions cause each component
of the zero
locus to contribute $+1$ or $-1$ to the path integral (according to
orientation). See~\cite{W:tqft} for an explanation of how this happens.
In this way, the path integral computes the {\em Euler class\/} of the
bundle $\W$. If $\dim\W=\dim\mm$, the Euler class is an integer. If
$\dim\W<\dim\mm$, the Euler class can be thought of as an homology
cycle within $\mm$. To obtain integers in this case, one can insert
cohomology classes (observables) into the path integral which leads to
intersection numbers on this cycle~\cite{W:tqft,W:tsm}.

It is instructive to do this calculation for the case of the single
cover of an isolated  rational curve~\cite{W:aspects}. The zero section of $s$
corresponds to the moduli space, $M$, of holomorphic maps of a $\CP1$ into
this curve, or equivalently, reparametrizations of $\CP1$. This space
is well-known to be $Sl(2,\SetCx)$. To do intersection theory, we
should really have a compact space, $\overline{M}$,
for the instanton moduli space. $Sl(2,\SetCx)$
naturally compactifies to $\CP3$. We can choose the coordinates $u^i$
to obtain the map
function
\begin{equation}
  u^3(z) = \frac{az+b}{cz+d}  \label{eq:sl2}
\end{equation}
and $u^1(z)=u^2(z)=0$. The complex parameters in (\ref{eq:sl2}) can be
chosen to satisfy $ad-bc=1$. This space naturally compactifies to $\CP3$
by considering the coordinates $[a,b,c,d\,]$ as homogeneous.
The space that we add in to $Sl(2,\SetCx)$ to form $\CP3$ (i.e., the
{\em compactification divisor\/}) is isomorphic to $\CP1\times\CP1$
and can be thought of as the algebraic variety defined by $ad-bc=0$.

To calculate (\ref{eq:11c}) we need to represent the $e_a$ as homology
classes within the compactified instanton moduli space,
$\overline{M}$. This can be achieved as follows~\cite{W:aspects}. Each $e_a$
is associated with a 4-cycle in $X$ or, by Poincar\'e duality, by a
(1,1)-form whose support lies in an arbitrarily small neighborhood
around this cycle. Thus any map which does not take a point of
$\Sigma$ to this cycle gives an arbitrarily small contribution to the
path integral. Using the same argument about the fermionic
determinants as we used above to obtain the Euler class, we can obtain
the following result:
\begin{equation}
  N^v_{abc}=\#(L_a\cap L_b\cap L_c)
\end{equation}
where $L_a$ is the subspace of $\overline{M}$ given by maps that take
a particular point on $\Sigma$ to a point within the 4-cycle
associated to $e_a$. Which particular point in $\Sigma$ is chosen for each
$e_a$ does not affect the above calculation because of the properties
of topological field theories.

For the case we are considering we get a contribution to $L_a$ for
each point of intersection between $C$ and $e_a$. Each such point
clearly puts a linear constraint on the homogeneous coordinates
$[a,b,c,d\,]$ and thus gives a hyperplane in $\overline{M}$. Three
hyperplanes in $\CP3$ intersect at one point and so we reproduce the
result
\begin{equation}
  N^{v(n=1)}_{abc}=\#(e_a\cap C).\#(e_b\cap C).\#(e_c\cap C).
	\label{eq:n-1}
\end{equation}

Now let us try to perform the same calculation for a multiple cover of
$C$. The additional complexity of this problem is caused by $\rho^i$
zero modes. The solution to this problem is provided by much the same
method as used in~\cite{W:matrx} where similar ``antighost'' zero
modes also appeared. To understand more clearly what the relationship
between the path integral and the Euler class is, we should consider
the approach of~\cite{AJ:}.

Consider a vector bundle $\W\to\mm$. We will refer to $\mm$ as the
horizontal direction of the bundle and the fibre as the vertical direction.
We can consider the cohomology of
this space by introducing differential forms with compact support in
the vertical direction, i.e., each form vanishes outside a compact
region of the fibre.
It is a general result of algebraic topology (see, for
example~\cite{BT:diff}) that there is a $(\dim\W)$-form in this class,
called the {\em Thom class\/} which can be pulled back via the zero
section to the Euler class on $\mm$. In the case we are considering we
have an bundle of infinite dimension. In this situation
it is more convenient to introduce the Mathai-Quillen
form~\cite{MQ:form},
\begin{equation}
  \omega_s=s^*U
\end{equation}
where $s$ is a section of $\W$ and $U$ is a slight variant of the Thom
class. One replaces the condition for compact vertical
support with the constraint that the form should decay at least as
fast as $\exp(-x^2)$ along the fibre. If $s$ is the zero section we
effectively reproduce the above argument whereas we can represent the
path integral idea if we take $s$ to be some $t\to\infty$ limit of a
generic section.

The moduli space $\mm$ in our theory is the space of {\em all\/} maps
$\Sigma\to X$.  It is convenient to assume that the maps under consideration
have derivatives in $L^p$; then using the Sobolev norm one can give $\mm$
the structure of a Banach manifold with an almost-complex structure
\cite{McD:,Klg:}.  With this structure, the holomorphic tangent space to
$\mm$ at $\phi$ is the space $T_{{\mm},\phi}:=H^0_{(p)}(\Sigma,\phi^*T_X)$
of $L^p$-sections of the bundle $R=\phi^*T_X$.  These spaces fit together
to form the holomorphic tangent bundle $T_{{\mm}}\to\mm$, which is a
Banach bundle over $\mm$.

For any map
$\phi:\Sigma\to X$, the derivative $d\phi$ gives a map between tangent
bundles $T_\Sigma\to T_X$, which can be thought of as a map
$T_\Sigma\to\phi^*T_X$.  If we let $\cal A^1_\Sigma$ denote the 1-forms
on $\Sigma$, $d\phi$ can also be regarded a section of
$\cal A^1_\Sigma\otimes\phi^*T_X$ (which would determine a map from $T_\Sigma$
by evaluating the 1-form on the tangent vector).  Taking the
(0,1)-part of $d\phi$, we get
$\bar\partial\phi$, which can be interpreted as a section of
$\cal A^{0,1}_\Sigma\otimes\phi^*T_X$ over $\Sigma$, that is, as an
element of the space
$\W_\phi:=H^0_{(p)}(\Sigma,\cal A^{0,1}_\Sigma\otimes\phi^*T_X)$.
The spaces $\W_\phi$
fit together to form another Banach bundle $\W\to\mm$.
This bundle comes equipped with the natural section
\begin{equation}
  s:\phi\mapsto \bar\partial\phi.
\end{equation}
This is the section $s$ which appears in the action (\ref{eq:act1}), and
whose zero locus $M$ is the space of holomorphic maps.

We can also consider the following more heuristic approach to the
bundle $\W$ in terms of local coordinates.
The fermions in the path integral can be taken to
represent differential forms on $\W$.\footnote{In topological
Yang-Mills there are three sets of fermions. The extra set arises
because of gauge invariance and is not relevant here.} To be more precise, the
``anti-ghost'' fermions $\rho^i$ represent a basis of 1-forms in the vertical
direction and the ``ghost'' fermions $\chi^i$ represent a basis of
1-forms in the horizontal direction. The following diagram should be
born in mind:
\begin{equation} \label{eq:cd}
  \begin{CD}
    u^i @>\bar\partial>> \bar\partial u^i  \\
     @V\delta VV           @VV\delta V  \\
    \delta u^i @>\bar D>> \bar D\delta u^i
  \end{CD}
\end{equation}
The bottom row of this diagram can be thought of as 1-forms with
$\delta$ acting as some kind of de Rham $d$-operator. The
left-hand side can be thought of as the horizontal part of the bundle
and the right-hand side the vertical part. The correspondence between
$\delta u^i$ and $\chi^i$ is central to topological quantum field
theory and is what allows us to represent BRST-observables as
cohomology classes in the moduli space~\cite{W:tqft}. In the vertical
direction the fibre should be thought of as $\SetRe^\infty$ and so the
cotangent bundle in this direction is isomorphic to the fibre itself.
Thus the $\rho^i$ fields can be interpreted as a basis for the fibre.
When we perform the path integral (\ref{eq:11c}) we can interpret this
as taking the Euler class of the vector bundle whose fibre is spanned
by the $\rho^i$ fields. That is, the vector bundle in question is the
bundle of $(0,1)$-forms on $\Sigma$ taking values in the pullback of
the tangent bundle on $X$.

The moduli space $\mm$ has been closely studied by Gromov \cite{Gro:}.
(In fact,
Gromov's work provided one of the inspirations for Witten's construction
of topological $\sigma$-models \cite{W:tsm}.)
We follow the approaches of McDuff \cite{McD:} and Wolfson \cite{Wol:}
to Gromov's ideas.  Gromov showed that $\mm$
has a natural compactification to a space $\overline{\mm}$ which includes
maps from ``simple cusp-curves'' $\Sigma_0 \cup\bigcup_{\alpha=1}^k \P^1$
to $X$ as limits of maps from $\Sigma\to X$.  In the case at hand (multiple
covers of an isolated rational curve $C$), these maps from
cusp-curves can be most
easily understood by means of their graphs.  A sequence of graphs
$\Gamma_{\phi_j}\subset\Sigma\times X$
of maps $\phi_j:\Sigma\to X$ can converge to a curve in $\Sigma\times X$
which is {\em not} a graph.  Such a limiting graph
$\Gamma_{\phi}$ will have one component $\Sigma_0$
which is the graph of a map of lower degree (possibly even degree 0),
and other $\P^1$-components which map to points in $X$.

If we extend $\W$ to a bundle $\overline{\W}$ over the compactified space
$\overline{\mm}$ by continuing to use the spaces
$\W_\phi:=H^0_{(p)}(\Sigma,\cal A^{0,1}_\Sigma\otimes\phi^*T_X)$
as fibres even in the case of limiting graphs, then the limiting maps
$\phi$ will still have $\bar\partial\phi\in\W_\phi$.  Thus, the section
$s$ extends to a section $\bar s$ of $\overline{\W}\to\overline{\mm}$.

We now attempt to calculate the Euler class by pulling back the
Mathai-Quillen form by our section $\bar s$ of
$\overline{\W}\to\overline{\mm}$.
For this process to work correctly, we are required to take a {\em
generic\/} section of this bundle.
Unfortunately, our section $\bar s$ is usually not generic.  Fortunately,
there is a method \cite{W:matrx} which can be used to calculate the
Euler class from a non-generic section.  (We have formally extended this
method from the finite-dimensional to the infinite-dimensional case,
ignoring convergence questions.)

If we vary the map $\phi$ by a displacement $\delta\phi$ (or in local
coordinates vary $u^i$ by $\delta u^i$ as in (\ref{eq:cd})), we find that
$\delta\phi\in T_{{\mm},\phi}=H^0_{(p)}(\Sigma,\phi^*T_X).$
Thus, the variation of $\bar\partial\phi$ is given by
\begin{equation}
\bar D \delta\phi\in\bar D  H^0_{(p)}(\Sigma,\phi^*T_X).
\end{equation}
If we regard $\bar D $ as defining a map of bundles
\begin{equation}
\bar D :T_{\overline{\mm}}\to\overline{\W}
\end{equation}
(a linearization of the section $\bar\partial$),
we find that the displacements of $\bar\partial\phi$ all lie in the image
bundle $\W^\prime=\Image(\bar D )$.

Following~\cite{W:matrx} we can use this information to calculate the
correct Euler class of $\overline{\W}$. We have the following exact sequence:
\begin{equation}
  0 @>>> \W^\prime @>i>> \overline{\W} @>>> \V @>>> 0.
\end{equation}
The cokernel $\V$ is a bundle
whose fibres correspond to $H^1(R)$. This bundle has finite dimension
$2n-2$. The zero locus of a generic section of $\overline{\W}$ can be taken as
the zero locus of a generic section of $\V$ restricted to the zero
locus of a generic section of $\W^\prime$. Translating this into a
statement about Euler classes, we can do the path integral
(\ref{eq:11c}) by considering the moduli space of holomorphic maps
(the Euler class of $\W^\prime$) and including the cohomology class
corresponding to the Euler class of $\V$ (the top Chern class
of $\V$) in the integrand. That is
\begin{equation}
  \eqalign{N^v_{abc}&=\int_{M_v}e_ae_be_c.c_{2n-2}(H^1(R))\cr
	&=\#(L_a\cap L_b\cap L_c\cap U),\cr} \label{eq:ms}
\end{equation}
where the $e_a$'s are interpreted as (1,1)-forms on $M_v$ and $U$ is a
6-cycle corresponding to the Poincar\'e dual of the Euler class of $\V$.

The appearance of the top Chern class in (\ref{eq:ms}) can be viewed
as the effect of integrating out the $\rho^i$ zero modes by using the
four-fermion term in the action (\ref{eq:11a}). This process will
bring powers of the curvature tensor into the action and this should
correspond to the usual de Rham representation of Chern classes by
powers of the curvature.

We have reduced the problem of finding the Euler class of an infinite
dimensional vector bundle, which is best done by path integral
techniques, to the problem of finding the Euler class of a finite
dimensional vector bundle. This can be done by more conventional
methods in algebraic geometry.


\section{A Bundle Calculation}

Let $X$ be a Calabi-Yau threefold, and let $C \subset X$ be an isolated smooth
rational curve such that $T_X|_{C}\cong \O_C(2)\oplus\O_C(-1)\oplus\O_C(-1)$.
Consider the moduli space
\begin{equation}
M_n(C)=\{\phi:\P^1\to X\ |\ \phi(\P^1)=C, \deg\phi=n\}
\end{equation}
of parametrized maps from $\Sigma=\P^1$ to $X$.
For each $\phi\in M_n(C)$,
the vector space $H^0(\phi^*(T_X))$ is the tangent space at $\phi$
to this moduli
problem,\footnote{Since $C$ is isolated in $X$, the conditions $\phi(\P^1)=C$,
$\deg\phi=n$ simply serve to pick out a component of the space of all maps
from $\P^1$ to $X$.}
while the vector space $H^1(\phi^*(T_X))$ is the obstruction
space (also at $\phi$).  The ``virtual
dimension'' of the moduli space is therefore
\begin{equation}
\dim H^0(\phi^*(T_X)) - \dim H^1(\phi^*(T_X)) = \chi(\phi^*(T_X))=3.
\end{equation}

Motivated by Gromov's work \cite{Gro:} as described in the previous
section, we compactify $M_n(C)$ by using graphs.
Associate to each $\phi\in M_n(C)$ the graph
\begin{equation}
\Gamma_{\phi}\subset\P^1\times C.
\end{equation}
These graphs fit together into a ``universal graph''
\begin{equation}
\Gamma\subset M_n(C)\times\P^1\times C
\end{equation}
defined by
\begin{equation}
\Gamma=\{(\phi,t,\phi(t))\ |\ \phi\in M_n(C), t\in\P^1\}.
\end{equation}
The graphs $\Gamma_{\phi}$ all belong to a common linear system
on $\P^1\times C$.  We compactify $M_n(C)$ to
$\overline M_n(C)\cong\P^{2n+1}$ by
including the elements of that linear system which are {\em not} graphs
of maps of degree $n$.  The universal graph then compactifies to the
``universal divisor in the linear system''
\begin{equation}
\overline\Gamma\subset\overline M_n(C)\times\P^1\times C.
\end{equation}
This has a very concrete description as follows.  Let $x$, $y$ be
homogeneous coordinates on $\P^1$, let $s$, $t$ be homogeneous coordinates
on $C$, and let $a_0,\dots,a_n,b_0,\dots,b_n$ be homogeneous coordinates
on $\overline M_n(C)$.  Then the map $\phi$ corresponding to the
point $[a_0,\dots,a_n,b_0,\dots,b_n]$ can be described as
\begin{equation}
\frac st=\frac{\sum a_ix^iy^{n-i}}{\sum b_ix^iy^{n-i}}
\end{equation}
and the divisor $\overline\Gamma$ in
$\overline M_n(C)\times\P^1\times C$ is defined by the equation
\begin{equation} \label{eq1}
\sum a_ix^iy^{n-i}t - \sum b_ix^iy^{n-i}s=0.
\end{equation}
As before, each new graph which has been added in this compactification
contains a graph of a map of lower degree, together with some $\P^1$'s
which map to points in $C$.

It is a simple matter to explicitly construct the compactification divisor of
$\overline M_n(C)$. The points in $\overline M_n(C)$ which do not
correspond to smooth $n$-fold covers are given by values of
$[a_0,\ldots,b_n]$ for which (\ref{eq1}) factorizes. That is, the
resultant of the two polynomials of (\ref{eq1}) vanishes:
\begin{equation}
  \begin{vmatrix}
     a_0&a_1&a_2&\ldots&0\\
     0&a_0&a_1&\ldots&0\\
     &&\vdots&&\\
     0&0&0&\ldots&a_n\\
     b_0&b_1&b_2&\ldots&0\\
     &&\vdots&&\\
     0&0&0&\ldots&b_n
  \end{vmatrix}
  =0.	  \label{eq:resnt}
\end{equation}
Thus, the compactification divisor is a hypersurface of degree $2n$
in $\overline M_n(C)\cong\CP{2n+1}$. This shows that $M_n(C)$ is
isomorphic to the subspace of $\SetCx^{2n+2}$ given by the constraint
that the left side
of (\ref{eq:resnt}) is equal to 1. This generalizes the $ad-bc=1$
constraint for the single cover case.

It will be convenient to denote $\P^1\times C$ by $S$.
Let $p:S\to\P^1$ and
$q:S\to C$ be the projection maps;
let $p_{\phi}:\Gamma_{\phi}\to\P^1$ and
$q_{\phi}:\Gamma_{\phi}\to C$
be the induced maps on the graphs.
If $\phi\in M_n(C)$, then $p_{\phi}$ establishes a natural isomorphism
between $\Gamma_{\phi}$ and $\P^1$.

We regard $T_X|_C$ as a fixed bundle on $C$.  Then for each $\phi\in
M_n(C)$, the bundle $\phi^*(T_X|_C)$ is mapped to the bundle
$q_{\phi}^*(T_X|_C)$ under the isomorphism $p_{\phi}$.  Thus, the
important spaces for us to study are the spaces $H^0(q_{\phi}^*(T_X|_C))$
and $H^1(q_{\phi}^*(T_X|_C))$.  (As $\phi$ varies, these spaces will
fit together to form
bundles over the moduli space $M_n(C)$.)
We now need to extend this bundle over $M_n(C)$ to a sheaf over the
compactified space $\overline M_n(C)$. Unfortunately this is not a
unique process. However, given the construction we have used for the
compactified moduli space, there is a natural way we can do this.
As in the previous section,
we choose to extend by using $H^0(q_{\phi}^*(T_X|_C))$
and $H^1(q_{\phi}^*(T_X|_C))$ for all values of $\phi$, even values
for which $\Gamma_{\phi}$ is not a graph.

We now claim that the dimensions of these spaces are independent of
$\phi$, and that they admit another description.
Let $\B=q^*(T_X|_C)$.  Then the restriction $\B|_{\Gamma_{\phi}}$
of $\B$ to $\Gamma_{\phi}$ coincides with $q_{\phi}^*(T_X|_C)$.
Thus, the spaces we wish to study can be identified with
$H^0(\B|_{\Gamma_{\phi}})$ and $H^1(\B|_{\Gamma_{\phi}})$.
Consider now the restriction sequence
\begin{equation} \label{eq2}
0\to\B(-\Gamma_{\phi})\to\B\to\B|_{\Gamma_{\phi}}\to0.
\end{equation}
The associated long exact sequence in cohomology is then
\begin{equation} \label{eq3}
{\arraycolsep 2pt
\begin{array}{ccccccccccc}
0 & \to &  H^0(\B(-\Gamma_{\phi})) & \to &  H^0(\B) & \to &
H^0(\B|_{\Gamma_{\phi}})  &
\to  & H^1(\B(-\Gamma_{\phi})) &  & \\
\hphantom{0} & \to  & H^1(\B) & \to &
H^1(\B|_{\Gamma_{\phi}})  & \to  & H^2(\B(-\Gamma_{\phi})) & \to  & H^2(\B) &
\to & 0.
\end{array}
}
\end{equation}
(The last $H^2$ term is 0 since $\supp(\Gamma_\phi)$ is a curve.)

We will abuse notation a bit and let $\P^1$ and $C$ denote divisor classes
on $S$, as follows:  for some fixed points $P\in C$, $Q\in \P^1$ we identify
$\P^1=q^{-1}(P)$, $C=p^{-1}(Q)$.  Since equation (\ref{eq1}) has
degree $1$ in $s$ and $t$, and degree $n$ in $x$ and $y$,
 each $\Gamma_{\phi}$ belongs to
the linear system $|\P^1+nC|$ on $S$.
Moreover, since $T_X|_{C}\cong \O_C(2)\oplus\O_C(-1)\oplus\O_C(-1)$ by
assumption, we have
\begin{equation}
\B=q^*(T_X|_C)\cong \O_S(2\P^1)\oplus\O_S(-\P^1)\oplus\O_S(-\P^1).
\end{equation}
It follows that
\begin{equation} \label{eq3.5}
\B(-\Gamma_{\phi})
\cong \O_S(\P^1-nC)\oplus\O_S(-2\P^1-nC)\oplus\O_S(-2\P^1-nC).
\end{equation}
Note also that $K_S=-2\P^1-2C$.

We calculate the cohomology of $\B$ as follows.  First,
\begin{equation}
H^0(\B)=H^0(\O_S(2\P^1)\oplus\O_S(-\P^1)\oplus\O_S(-\P^1))\cong
H^0(\O_S(2\P^1))
\end{equation}
which has dimension 3.  Also, by Serre duality,
\begin{equation}
H^2(\B)^*\cong
H^0(\O_S(-4\P^1-2C)\oplus\O_S(-\P^1-2C)\oplus\O_S(-\P^1-2C))=\{0\}.
\end{equation}
In addition, since $c_1(\B)=c_2(\B)=0$, by Riemann-Roch we have $\chi(\B)=3.$
It follows that $H^1(\B)=\{0\}$ as well.

Next we calculate the cohomology of $\B(-\Gamma_{\phi})$.
It follows directly from equation (\ref{eq3.5}) that
$H^0(\B(-\Gamma_{\phi}))=0$, while by Serre duality
\begin{eqnarray*}
H^2(\B(-\Gamma_\phi))^* & \cong &
H^0(\O_S(-3\P^1+(n-2)C)\oplus\O_S((n-2)C)\oplus\O_S((n-2)C)) \\
 & \cong &
H^0(\O_S((n-2)C)\oplus H^0(\O_S((n-2)C),
\end{eqnarray*}
which has dimension $(n-1)+(n-1)=2n-2$.
This time, $c_1(\B(-\Gamma_{\phi}))=-3\P^1-3nC$ and
$c_2(\B(-\Gamma_{\phi}))=6n$.  Thus, by
Riemann-Roch, $\chi(\B(-\Gamma_{\phi}))=0$.  It
follows that $H^1(\B(-\Gamma_{\phi}))$ also has dimension $2n-2$.

Thus we see that the long exact sequence (\ref{eq3}) can be
shortened to
\begin{equation}
 0  \to  H^0(\B)  \to  H^0(\B|_{\Gamma_{\phi}})
   \to  H^1(\B(-\Gamma_{\phi}))  \to  0  \to
    H^1(\B|_{\Gamma_{\phi}})
  \to  H^2(\B(-\Gamma_{\phi}))  \to  0.
\end{equation}
Moreover, the dimensions of all of these spaces are independent of
$\phi$, so the sheaf we are after will be locally free.

Now to fit these all together, let $\overline M=\overline M_n(C)$, and
look at the space $Z=\overline M\times S$,
with projections $\pi:Z\to\overline M$, $\rho:Z\to S$.  Let
$\E=\rho^*(\B)$.
The restriction sequences (\ref{eq2}) fit together into a restriction sequence
\begin{equation} \label{eq4}
0\to\E(-\overline\Gamma)\to\E\to\E|_{\overline\Gamma}\to0.
\end{equation}
For any bundle ${\cal G}$ on $Z$, $R^i\pi_*\cal G$ denotes the bundle
of $H^i(\text{fibre},\cal G_{\text{fibre}})$'s on $\overline M$.
Using the vanishing of cohomologies which we have established above,
we see that
the long exact sequence associated to (\ref{eq4}) takes the form
\begin{equation}
0\to R^0\pi_*\E\to R^0\pi_*(\E|_{\overline\Gamma})\to
R^1\pi_*(\E(-\overline\Gamma))\to0\to R^1\pi_*(\E|_{\overline\Gamma})\to
R^2\pi_*(\E(-\overline\Gamma))\to0.
\end{equation}
In particular, $R^1\pi_*(\E|_{\overline\Gamma})$
is isomorphic to $R^2\pi_*(\E(-\overline\Gamma))$, and
both of these are locally free of rank $2n-2$.

To identify the bundle $R^2\pi_*(\E(-\overline\Gamma))$, we need
one additional fact:
$\overline\Gamma\in|\pi^{-1}(H)+\rho^{-1}(\P^1)+n\rho^{-1}(C)|$,
where $H$ is a hyperplane in $\overline M\cong\P^{2n+1}$.  This holds because
the equation for $\overline\Gamma$, equation (\ref{eq1}), has degree $1$
in $a_0,\dots,b_n$, degree $1$ in $s,t$ and degree $n$ in $x,y$.

It follows that
 $\E(\pi^{-1}(H)-\overline\Gamma)$ will be a pullback from $S$.  That
is, there is some bundle $\F$ on $S$ such that
$\E(\pi^{-1}(H)-\overline\Gamma)=\rho^*(\F)$.  Then by the projection formula,
\begin{equation}
R^2\pi_*(\E(-\overline\Gamma))\cong
\O_{\overline M}(-H)\otimes R^2\pi_*(\rho^*(\F)).
\end{equation}
Now $R^2\pi_*(\E(-\overline\Gamma))$ and $R^2\pi_*(\rho^*(\F))$ are both
locally free of rank $2n-2$, and since $Z$ is the product of $\overline M$
and $S$, $R^2\pi_*(\rho^*(\F))$ must actually be the trivial bundle of
that rank.  Thus,
\begin{equation}
R^2\pi_*(\E(-\overline\Gamma))\cong
\O_{\overline M}(-H)\otimes ({\O_{\overline M}}^{\oplus (2n-2)})\cong
(\O_{\overline M}(-H))^{\oplus (2n-2)}. \label{eq:r2}
\end{equation}
The Hodge numbers for $\CP{2n+1}$ satisfy $h^{i,j}=\delta_{i,j}$ and
so we have a unique integral generator, $I$, for
$H^{2n-2,2n-2}(\CP{2n+1})$. Equation (\ref{eq:r2}) tells us that
\begin{equation}
  c_{2n-2}(R^1\pi_*(\E|_{\overline\Gamma}))=(-1)^{2n-2}I=I.
\end{equation}


\section{The Result}

We are now in a position to complete the computation of (\ref{eq:ms}).
The result of the previous section is that $U$ is homological to a
sub-$\CP3$ of the moduli space $\CP{2n+1}$. The argument now proceeds
in an identical way to the case of $n=1$. For each point of
intersection between $C$ and $e_a$ we obtain a contribution to $L_a$.
The condition that a point of $\Sigma$ maps into a specific point of
$X$ puts a linear constraint on the coordinates $[a_0,\ldots,b_n]$ of
$\overline M_n(C)$ and thus corresponds to a hyperplane of $\overline
M_n(C)$. The intersection of 3 hyperplanes and a sub-$\CP3$ within
$\CP{2n+1}$ is a point and so we yield a result identical to
(\ref{eq:n-1}) for any $n>0$
\begin{equation}
  N^{v}_{abc}=\#(e_a\cap C).\#(e_b\cap C).\#(e_c\cap C).
	\label{eq:n-n}
\end{equation}

As an example, consider the case examined in~\cite{Cand:rat}. We take
$X$ to be the algebraic variety given by a quintic constraint in
$\CP4$. This manifold has the simplifying feature that $h^{1,1}=1$. In
this case we only have one $f$-polynomial of the form (\ref{eq:fab})
and this polynomial is a function of only one variable, $q$.

Let us, for the time being, make the assumption that all rational
curves are of the type ($-1$,$-1$).
In the case of $h^{1,1}=1$, each rational curve in $X$ has a {\em
degree\/} which is defined by the intersection number for this curve
with the unique integral generator of $H^{1,1}(X)$. For the quintic
threefold this generator is the generator inherited from the ambient
$\CP4$.  Let us compute the
contribution to the $f$-polynomial by an $n$-fold cover of a degree
$k$ curve. As we pullback the \K\ form from $X$ to $\Sigma$ by this
map we multiply the degree of this form on $\Sigma$ by $n$ and $k$. It
is thus a simple matter to show that $m=nk$ in (\ref{eq:scm}). From
(\ref{eq:n-n}) we see immediately that the contribution to $f$ is
\begin{equation}
  k^3q^{nk}. \label{eq:result}
\end{equation}
Thus
\begin{equation}
  f = 5+\sum_{k=1}^\infty\frac{a_kk^3q^k}{1-q^k},
\end{equation}
where $a_k$ is the number of rational curves of degree $k$ on $X$.
This is precisely the formula used in~\cite{Cand:rat} where the
expression (\ref{eq:result}) was conjectured and
justified by the fact that it was the
only simple form that would yield integers for $a_k$.

One of the great virtues of the $f$-polynomials comes from the mirror
property. The $f$-polynomials here are written in terms of the
complexified \K\ form and give information about the couplings between
(1,1)-forms on $X$. However, they also apply to couplings between
(2,1)-forms on the mirror of $X$ where here the $q_i$ variables would encode
information about the complex structure. This is how $f$ was derived
in~\cite{Cand:rat}. One key point to note is that the $f$-polynomials
were derived in the large-radius, or small $q$, limit. When using the
$f$-polynomials in the context of complex structure one must ensure
that one is in some ``large complex structure'' limit~\cite{AL:qag}. There
is a natural construction of this situation using methods of algebraic
geometry~\cite{Mor:math}.

The mirror symmetry also came in to play when we derived our
topological field theory from the original N=2 superconformal field
theory. We chose the BRST symmetry to be generated by
(\ref{eq:N=2(1)}). If we had chosen this symmetry to be generated
by (\ref{eq:N=2(2)}) we would have obtained identical results except for the
fact that instantons would then have been given by antiholomorphic maps. We
could also try to work with a BRST generator given by the right hand
set of equations in (\ref{eq:N=2(1)}) and the right hand set of
(\ref{eq:N=2(2)}). This would have the result of leaving us with a
quite different topological quantum
field theory containing only the (2,1)-forms from
the original theory rather than the (1,1)-sector. The $f$-polynomials
generated would be the ones describing the complex structure
deformations. By this method we obtain the general result that the
coefficients in these polynomials must always be integers. This fact
is not apparent from conventional algebraic geometry.

So far, in our example of the quintic threefold, we assumed that the
rational curves were isolated. This is believed to be true for a
generic quintic threefold \cite{C:ajmaps}.
If we consider the Fermat quintic, i.e.,
the threefold given by the constraint
\begin{equation}
  z_0^5+z_1^5+z_2^5+z_3^5+z_4^5=0
\end{equation}
in $\CP4$ with homogeneous coordinates $[z_0,\ldots,z_4]$, the
set of curves of degree one consists of a continuous family that forms
a set of 50 cones~\cite{Katz:50}. A generic member of this set is of
type ($-3$,1). One might suspect that the form of $f$ will change in
this situation. However, when one computes $f$ from the mirror
manifold, no account of the \K\ class of the mirror of $X$ is taken.
Thus, on $X$, $f$ must be invariant under deformations of complex
structure --- $f$ is a local diffeomorphism invariant. This establishes that
if one did the path integral over these families of rational curves in
the Fermat case, one should obtain the same $f$-polynomial as in the
generic case of isolated curves. It would be interesting the confirm
this explicitly.

\section*{Acknowledgments}

It is a pleasure to thank S. Katz, C.A. L\"utken and E. Witten for
sharing invaluable insights into this work.
The work of D.R.M.\ was supported in part by NSF grant DMS-9103827.

\end{document}